\documentstyle[twoside,psfig]{article}
\catcode`\@=11
\long\def\@makefntext#1{
\protect\noindent \hbox to 3.2pt {\hskip-.9pt
$^{{\eightrm\@thefnmark}}$\hfil}#1\hfill}		

\def\@makefnmark{\hbox to 0pt{$^{\@thefnmark}$\hss}}	

\def\ps@myheadings{\let\@mkboth\@gobbletwo
\def\@oddhead{\hbox{}
\rightmark\hfil\eightrm\thepage}
\def\@oddfoot{}\def\@evenhead{\eightrm\thepage\hfil
\leftmark\hbox{}}\def\@evenfoot{}
\def\sectionmark##1{}\def\subsectionmark##1{}}

\oddsidemargin=\evensidemargin
\newcounter{sectionc}\newcounter{subsectionc}\newcounter{subsubsectionc}
\renewcommand{\section}[1] {\vspace{12pt}\addtocounter{sectionc}{1}
\setcounter{subsectionc}{0}\setcounter{subsubsectionc}{0}\noindent
	{\tenbf\thesectionc. #1}\par\vspace{5pt}}
\renewcommand{\subsection}[1] {\vspace{12pt}\addtocounter{subsectionc}{1}
	\setcounter{subsubsectionc}{0}\noindent
	{\bf\thesectionc.\thesubsectionc. {\kern1pt \bfit #1}}\par\vspace{5pt}}
\renewcommand{\subsubsection}[1] {\vspace{12pt}\addtocounter{subsubsectionc}{1}
	\noindent{\tenrm\thesectionc.\thesubsectionc.\thesubsubsectionc.
	{\kern1pt \tenit #1}}\par\vspace{5pt}}
\newcommand{\nonumsection}[1] {\vspace{12pt}\noindent{\tenbf #1}
	\par\vspace{5pt}}
\newcounter{appendixc}
\newcounter{subappendixc}[appendixc]
\newcounter{subsubappendixc}[subappendixc]
\renewcommand{\thesubappendixc}{\Alph{appendixc}.\arabic{subappendixc}}
\renewcommand{\thesubsubappendixc}
	{\Alph{appendixc}.\arabic{subappendixc}.\arabic{subsubappendixc}}

\renewcommand{\appendix}[1] {\vspace{12pt}
        \refstepcounter{appendixc}
        \setcounter{figure}{0}
        \setcounter{table}{0}
        \setcounter{lemma}{0}
        \setcounter{theorem}{0}
        \setcounter{corollary}{0}
        \setcounter{definition}{0}
        \setcounter{equation}{0}
        \renewcommand{\thefigure}{\Alph{appendixc}.\arabic{figure}}
        \renewcommand{\thetable}{\Alph{appendixc}.\arabic{table}}
        \renewcommand{\theappendixc}{\Alph{appendixc}}
        \renewcommand{\thelemma}{\Alph{appendixc}.\arabic{lemma}}
        \renewcommand{\thetheorem}{\Alph{appendixc}.\arabic{theorem}}
        \renewcommand{\thedefinition}{\Alph{appendixc}.\arabic{definition}}
        \renewcommand{\thecorollary}{\Alph{appendixc}.\arabic{corollary}}
        \noindent{\tenbf Appendix \theappendixc #1}\par\vspace{5pt}}
\newcommand{\subappendix}[1] {\vspace{12pt}
        \refstepcounter{subappendixc}
        \noindent{\bf Appendix \thesubappendixc. {\kern1pt \bfit #1}}
	\par\vspace{5pt}}
\newcommand{\subsubappendix}[1] {\vspace{12pt}
        \refstepcounter{subsubappendixc}
        \noindent{\rm Appendix \thesubsubappendixc. {\kern1pt \tenit #1}}
	\par\vspace{5pt}}

\newcommand{\textlineskip}{\baselineskip=13pt}
\newcommand{\smalllineskip}{\baselineskip=10pt}

\def\eightcirc{
\begin{picture}(0,0)
\put(4.4,1.8){\circle{6.5}}
\end{picture}}
\def\eightcopyright{\eightcirc\kern2.7pt\hbox{\eightrm c}}

\newcommand{\copyrightheading}[1]
	{\vspace*{-2.5cm}\smalllineskip{\flushleft
	{\footnotesize International Journal of Modern Physics C #1}\\
	{\footnotesize $\eightcopyright$\, World Scientific Publishing
	 Company}\\
	 }}

\newcommand{\publisher}[2]{{\begin{center}\footnotesize\smalllineskip
	Received #1\\
	Revised #2
	\end{center}
	}}

\def\abstracts#1#2#3{{
	\centering{\begin{minipage}{4.5in}\footnotesize\baselineskip=10pt
	\parindent=0pt #1\par
	\parindent=15pt #2\par
	\parindent=15pt #3
	\end{minipage}}\par}}

\def\keywords#1{{
	\centering{\begin{minipage}{4.5in}\footnotesize\baselineskip=10pt
	{\footnotesize\it Keywords}\/: #1
	\end{minipage}}\par}}

\renewenvironment{thebibliography}[1]
        {\frenchspacing
	 \ninerm\baselineskip=11pt
         \begin{list}{\arabic{enumi}.}
        {\usecounter{enumi}\setlength{\parsep}{0pt}
	 \setlength{\leftmargin 12.7pt}{\rightmargin 0pt} 
         \setlength{\itemsep}{0pt} \settowidth
	{\labelwidth}{#1.}\sloppy}}{\end{list}}

\newcounter{itemlistc}
\newcounter{romanlistc}
\newcounter{alphlistc}
\newcounter{arabiclistc}

\newcommand{\fcaption}[1]{
        \refstepcounter{figure}
        \setbox\@tempboxa = \hbox{\footnotesize Fig.~\thefigure. #1}
        \ifdim \wd\@tempboxa > 5in
           {\begin{center}
        \parbox{5in}{\footnotesize\smalllineskip Fig.~\thefigure. #1}
            \end{center}}
        \else
             {\begin{center}
             {\footnotesize Fig.~\thefigure. #1}
              \end{center}}
        \fi}

\newcommand{\tcaption}[1]{
        \refstepcounter{table}
        \setbox\@tempboxa = \hbox{\footnotesize Table~\thetable. #1}
        \ifdim \wd\@tempboxa > 5in
           {\begin{center}
        \parbox{5in}{\footnotesize\smalllineskip Table~\thetable. #1}
            \end{center}}
        \else
             {\begin{center}
             {\footnotesize Table~\thetable. #1}
              \end{center}}
        \fi}

\def\@citex[#1]#2{\if@filesw\immediate\write\@auxout
	{\string\citation{#2}}\fi
\def\@citea{}\@cite{\@for\@citeb:=#2\do
	{\@citea\def\@citea{,}\@ifundefined
	{b@\@citeb}{{\bf ?}\@warning
	{Citation `\@citeb' on page \thepage \space undefined}}
	{\csname b@\@citeb\endcsname}}}{#1}}

\newif\if@cghi
\def\cite{\@cghitrue\@ifnextchar [{\@tempswatrue
	\@citex}{\@tempswafalse\@citex[]}}
\def\citelow{\@cghifalse\@ifnextchar [{\@tempswatrue
	\@citex}{\@tempswafalse\@citex[]}}
\def\@cite#1#2{{$\null^{#1}$\if@tempswa\typeout
	{IJCGA warning: optional citation argument
	ignored: `#2'} \fi}}

\def\pmb#1{\setbox0=\hbox{#1}
	\kern-.025em\copy0\kern-\wd0
	\kern.05em\copy0\kern-\wd0
	\kern-.025em\raise.0433em\box0}

\def\fnt#1#2{\footnotetext{\kern-.3em
	{$^{\mbox{\scriptsize #1}}$}{#2}}}

\def\ps@myheadings{%
    \let\@oddfoot\@empty\let\@evenfoot\@empty
    \def\@evenhead{\slshape\leftmark\hfil}
    \def\@oddhead{\hfil{\slshape\rightmark}}
    \let\@mkboth\@gobbletwo
    \let\sectionmark\@gobble
    \let\subsectionmark\@gobble
    }
\font\tenrm=cmr10
\font\tenit=cmti10
\font\tenbf=cmbx10
\font\bfit=cmbxti10 at 10pt
\font\ninerm=cmr9

\font\eightrm=cmr8

\def\qed{\hbox{${\vcenter{\vbox{		    
   \hrule height 0.4pt\hbox{\vrule width 0.4pt height 6pt
   \kern5pt\vrule width 0.4pt}\hrule height 0.4pt}}}$}}

\def\bsc{{\sc a\kern-6.4pt\sc a\kern-6.4pt\sc a}}  	
\def\bflatex{\bf L\kern-.30em\raise.3ex\hbox{\bsc}\kern-.14em
T\kern-.1667em\lower.7ex\hbox{E}\kern-.125em X}

\begin{document}
\setlength{\textheight}{7.7truein}  

\thispagestyle{empty}

\markboth{\protect{\footnotesize\it T.Yamano}}
{\protect{\footnotesize\it Bornholdt's spin model of a market 
dynamics in high dimensions}}

\normalsize\textlineskip

\setcounter{page}{1}

\copyrightheading{}			

\vspace*{0.88truein}

\centerline{\bf Bornholdt's spin model of a market}
\vspace*{0.035truein}
\centerline{\bf dynamics in high dimensions}
\vspace*{0.37truein}
\centerline{\footnotesize Takuya Yamano\footnote{On leave from Department of
Applied Physics, Faculty of Science, Tokyo Institute of Technology,
Oh-okayama, Meguro-ku Tokyo, 152-8551, Japan\\
\it E-mail: tyamano@mikan.ap.titech.ac.jp}}
\baselineskip=12pt
\centerline{\footnotesize\it Institute for Theoretical Physics,
Cologne University, D-50923 K\"{o}ln, Euroland}
\centerline{\footnotesize\it E-mail: ty@thp.Uni-Koeln.DE}

\vspace*{0.225truein}
\publisher{(received date)}{(revised date)}

\vspace*{0.25truein}
\abstracts{.}{}{}
We present results of an extension of the market model introduced by 
Bornholdt to high dimensions. Three and four dimensions are shown to 
behave similar to two, for suitable parameters. 
\vspace*{5pt}


\vspace*{1pt}\textlineskip	
\section{Introduction}		
\vspace*{-0.5pt}
\noindent
To construct a reasonable dynamical financial market model consistent 
with the observed properties in real market is a central theme. Many 
models have been proposed along this line. Specifically application of 
concepts and techniques of statistical mechanics are giving deeper insight 
into the understanding of the complex behavior of a market.  
Fluctuations of prices of commodities, stocks and foreign exchange rates 
are examples of these. Reproduction of the distribution of returns based 
on the empirical data are a strong criterion to select models.  
It is important to set a communication structure for the model of financial 
market. Various models are proposed to imitate the real market communication. 
Among them we mention the minority game\cite{MG}, percolation\cite{ConB,perco} 
, Ising model\cite{Crem,Iori,Chow,Roeh,Kai,Born} and so forth. 
In this paper we focus on a 
spin model that seems to be appropriate to capture the movements of 
constituent agents in a simplified manner. In previous work, 
Iori\cite{Iori} has modeled three possible states of each agent's decision 
as spin-$1$ model and its decision making is affected by two different 
kind of noises and their nearest neighbors.\cite{Iori} However in this 
model the influence of market prices on each trader decision is not 
incorporated. Bornholdt recently formulated these influences with two 
different scales of local interaction and global one in two 
dimension.\cite{Born}
A metastable phase appears at randomly frozen finite magnetization which 
correspond to a bubble-like state in terms of economics.\cite{Born}\\
We assume that the traders live on a d-dimensional lattice. 
This totally contrasts the case\cite{Chow} where the 
{\it super-spins} do not need any geometry. 
In this paper we present the results of higher dimensions in the Bornholdt 
model. This can be motivated 
from the fact that one of the most important characters of the market 
evolution is the heterogeneous structure of agents. The heterogeneity arises 
from not only the geometrical distance of agents but also the amount of 
knowledge that each agents has, accessibility of market 
information prevailing, preferences, and processing skills for investments 
etc. 
  
\section{The Bornholdt Model}
\vspace*{-0.5pt}
\noindent
In the model, the spin variables are either $+1$ or $-1$, which allows
each agent to decide actions of buying or selling at each time step $t$. 
Thus the magnetizetion $M(t)=\sum_{j=1}^NS_j(t)/N$ can be interpreted as 
a measure of price.\cite{Born} We identify the logarithm of the absolute 
value of the magnetization as {\it returns}, i.e. $ret(t)=\ln | M(t)|-
\ln | M(t-1)|$. The use of linear returns (or simply relative price change) 
is also possible and in fact the main properties are also 
seen for this definition.\cite{Bornp} Each spin is updated by the 
following heat bath dynamics, 
\begin{equation}
S_i(t+1)=
\left\{
\begin{array}{rl}
  +1 & \mbox{with $p=[1+\exp(-2\beta h_i(t))]^{-1}$}\\
  -1 & \mbox{with $1-p$}
\end{array}
\right.,
\end{equation}
where $\beta$ is the inverse temperature and $h_i(t)$ is a time-dependent 
local field that each spin feels. This corresponds to a {\it signal} that 
each agent $i$ receives at time $t$\cite{Iori} and also to 
''individual bias'' of the $i$-th agent.\cite{Chow} The {\it temperature} is 
introduced in a totally parallel way to the Ising spin model in magnetism. 
This fictitious temperature has been used in the previous works.
\cite{Chow,Roeh} 

Bornholdt introduced the following form of the local field,
\begin{equation}
h_i(t)=\sum_{j=i}^NJ_{ij}S_j(t)-\alpha S_i(t)\Big | \frac{1}{N}\sum_{j=1}
^NS_j(t)\Big|\label{eqn:red}
\end{equation}
The above form can be considered incorporating two contradictory movements 
of constituent agents in a real market: herd behavior and preference of 
minority. $J_{ij}$ denotes the strength of the communication between the 
agents $i$ and $j$. Furthermore $J_{ij}$ is set to a constant $J$ for nearest 
neighbor agents and zero for all the other pairs. This term corresponds 
to the ``disagreement function'' of the $i$-th agent.\cite{Chow} $\alpha$ 
is a strength of the coupling to the magnetization and assumed to be 
positive. In a spin language, the first term prefers the ferromagnetic state. 
The second term, on the other hand, tends to encourage a spin flip when 
magnetizations becomes large.\cite{Born}\\

The more complicated form of the local field is given as\cite{Born} 
\begin{equation}
h_i(t)=\sum_{j=i}^NJ_{ij}S_j(t)-\alpha C_i(t)\frac{1}{N}\sum_{j=1}^NS_j(t)
\label{eqn:org},
\end{equation}
meaning we have two spins on each site at each time: demand spin $S_i(t)$ 
and strategy spin $C_i(t)$ of an agent $i$. Consideration in a market 
context gives an interpretation of strategy spin as\cite{Born}  
\begin{equation}
C_i(t)=
\left\{
\begin{array}{rl}
  +1 & \mbox{{\rm anti-ferro (fundamentalist)}}\\
  -1 & \mbox{{\rm ferro (chartist)}}
\end{array}
\right.
\end{equation}
This reminds us of the Lux-Marchesi model\cite{Lux}, where agents are divided 
into these two groups and evolve with interacting mutually and moreover 
with changing their strategies. We flip the strategy spin at next time step: 
$C_i(t+1)=-C_i(t)$ if the $S_i(t)$ is antiparallel to the global coupling 
to the magnetization i.e.,
\begin{equation}
S_i(t)C_i(t)\frac{1}{N}\sum_{j=1}^NS_j(t)<0
\end{equation}
is satisfied at each time because a positive energy contribution (risky or 
preference being minority) to the dynamics encourages these fundamentalists 
to switch their strategies to chartists and vice versa.
These updates are done after updating $S_i(t)$.

\section{Simulations and Results}
\vspace*{-0.5pt}
\noindent
We performed the simulation with the reduced version of the local field 
eq.(\ref{eqn:red}) 
except for dynamical evolutions of the fundamentalists-chartists behavior. 
The same behavior is observed either with eq.(\ref{eqn:org}) or with 
eq.(\ref{eqn:red}). 
The size of the hyper cubic lattice was set at $L=101$, $21$, and $7$ 
for the dimension $d=2$, $3$ and $4$ respectively. We have chosen the 
temperature (in units of $J/k_B$) $T=1.5$, $4.0$ and $6.0$ for the 
dimension $d=2$, $3$ and $4$ 
respectively. Note that these temperature is below the critical temperature 
$T_c=2.269 (2d)$, $4.511 (3d)$ and $6.680 (4d)$ in the case of $\alpha =0$ .
This selection of parameters allows us to observe the intermittent behavior 
in temporal evolution of $ret(t)$ or clustered volatilities. 
Fig.1 shows the distribution of returns for three different dimensions, 
whose shapes are non-Gaussian. The volatility of the returns or the 
generalized cumulative absolute returns\cite{Pasq} can be measured with 
the following quantity\cite{Liu,Pasq}
\begin{equation}
V(t)=\frac{1}{n}\sum_{t'=t}^{t+n-1}\big | ret(t')\big|^{\gamma},
\end{equation}
where $n$ is the number of data sampled and $\gamma$ is a real exponent. 
We set $n=1$ and $\gamma =1$, which leads to the cumulative distribution 
of absolute returns (Fig.2).  The measured exponents of these curves are 
$-1.15$, $-1.41$ and $-1.50$ respectively using the intermediate region 
of $8\dots 150$, $15\dots 150$ and $40\dots 150$ for fitting, where power-law 
scaling can be seen.  A volatility clustering 
can be seen as a slow decay of the autocorrelation function of 
absolute returns in Fig.3 on a log-log plot. Fig4. is the replot of the same 
data obtained in our simulation. The slope of the each curves 
is determined as $-1.85\times 10^{-3}$, $-1.80\times 10^{-2}$ and 
$-7.90\times 10^{-2}$ with the fitting time $350\dots 1000 (2d)$, 
$40\dots 140(3d)$ and $5\dots 35(4d)$ respectively, 
where the accurate value changes as the fitting regions vary. 
Fig.4 shows the ratio of the difference between the number of 
fundamentalists and that of chartists to the total number of agents with 
eq(\ref{eqn:org}) in the 4d case. The decrease in the ratio of 
fundamentalists corresponds to high variance in changes of returns.  

\section{Discussion and Summary}
\vspace*{-0.5pt}
\noindent
In this paper, we have performed simulations of the recently proposed model 
by Bornholdt in hyper cubic lattices. We confirmed that there is 
no intermittent behavior in return(magnetization) when we employ only 
the second term of eq.(\ref{eqn:red}). This means that market intermittency 
emerge as a consequence of a frustration across two different scales: local 
connection with neighbours and global coupling with market.\cite{Born}
The effects of dimensionality was confirmed, which exhibits similar 
behavior for three different cases. 
Moreover we confirmed that the similar intermittent behavior in returns 
appear when we drop out $S_i$ in the second term of the local field 
(eq.(\ref{eqn:red})). In this case, however, the fictitious temperature of 
the market becomes higher than that of the original one($T= 7.5$, $9.3$ 
and $11.0$ in the $2d$, $3d$, and $4d$ case respectively with the same 
other parameter values as in the text). However their temperatures exceed 
the critical ones. 
The exponent of cumulative distribution of returns is one of the ongoing 
discussion points for many models.\cite{perco,Biham} 
The recent measurement of the exponent based on a empirical data 
(S\&P $500$)\cite{Gopi} says the asymptotic slope of close to $-3$. 
A percolation model provides $-2.9$ 
with $201\times 201$ lattice\cite{perco} and $-4$ in high dimensions. 
In this sense, percolation model gives reasonable distribution. 
On the one hand, the model of Biham {\it et al.} provided the 
range of $-2.5 \dots -3.5$ for the tails of the distribution,\cite{Biham} 
which depends on the number of agents and a factor characterizing the 
market dynamics.
The slope (moreover the shape) of the cumulative distribution is 
strongly affected by the temperature in the present model. 
We confirmed the effect in the 3d case by fixing the other parameters. 
This means the power-law scaling in intermediate regions is very sensitive 
to the temperature. More specifically, cumulative distributions have a 
shoulder when the temperature is not the {\it tuned} value.  
The intermittency  
becomes more sparse as the temperature descends. The same can be seen as the 
lattice size becomes large. As for autocorrelation function,
the empirical data are well fitted with the exponent $-0.3$ according to 
Gopikrishnan et al.\cite{Gopi} and Liu et al.\cite{Liu}.
The exponent determined from minority game\cite{MG} is $-0.64$. 
In the present model, the tails of the autocorrelation 
function do not behave like power laws but as exponential up to four 
dimensions. 

\noindent
\begin{figure}[htbp]
\vspace*{13pt}
\centerline{\psfig{file=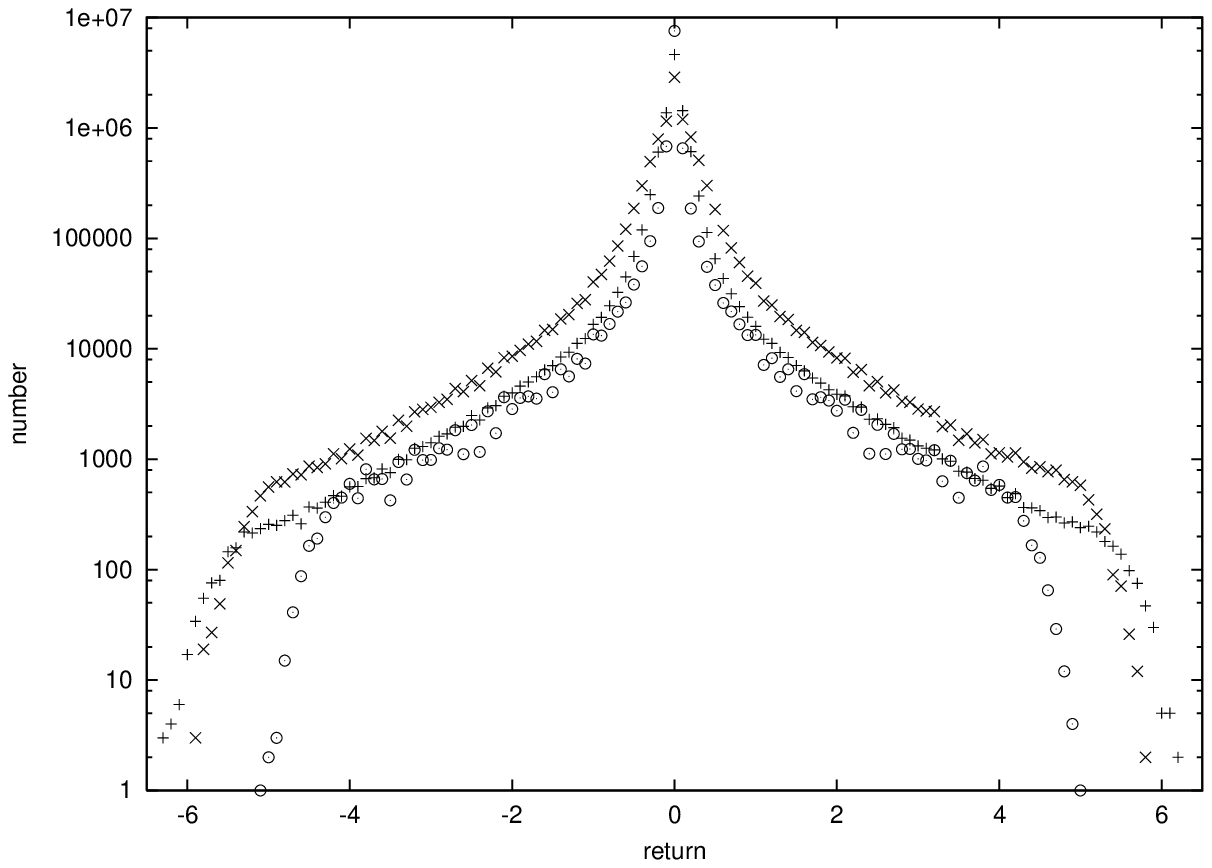}} 
\vspace*{13pt}
\fcaption{Distribution of the return defined as the difference 
of the logarithm of the absolute magnetization. 
Parameters $J=1.0$ and $\alpha =4.0$ are 
common for all dimensions. The symbols $\odot$, $+$ and $\times$ 
correspond to $(d,T)=(2,1.5)$, $(3,4.0)$, and $(4,6.0)$ respectively.}
\end{figure}

\noindent  
\begin{figure}[htbp]
\vspace*{13pt}
\centerline{\psfig{file=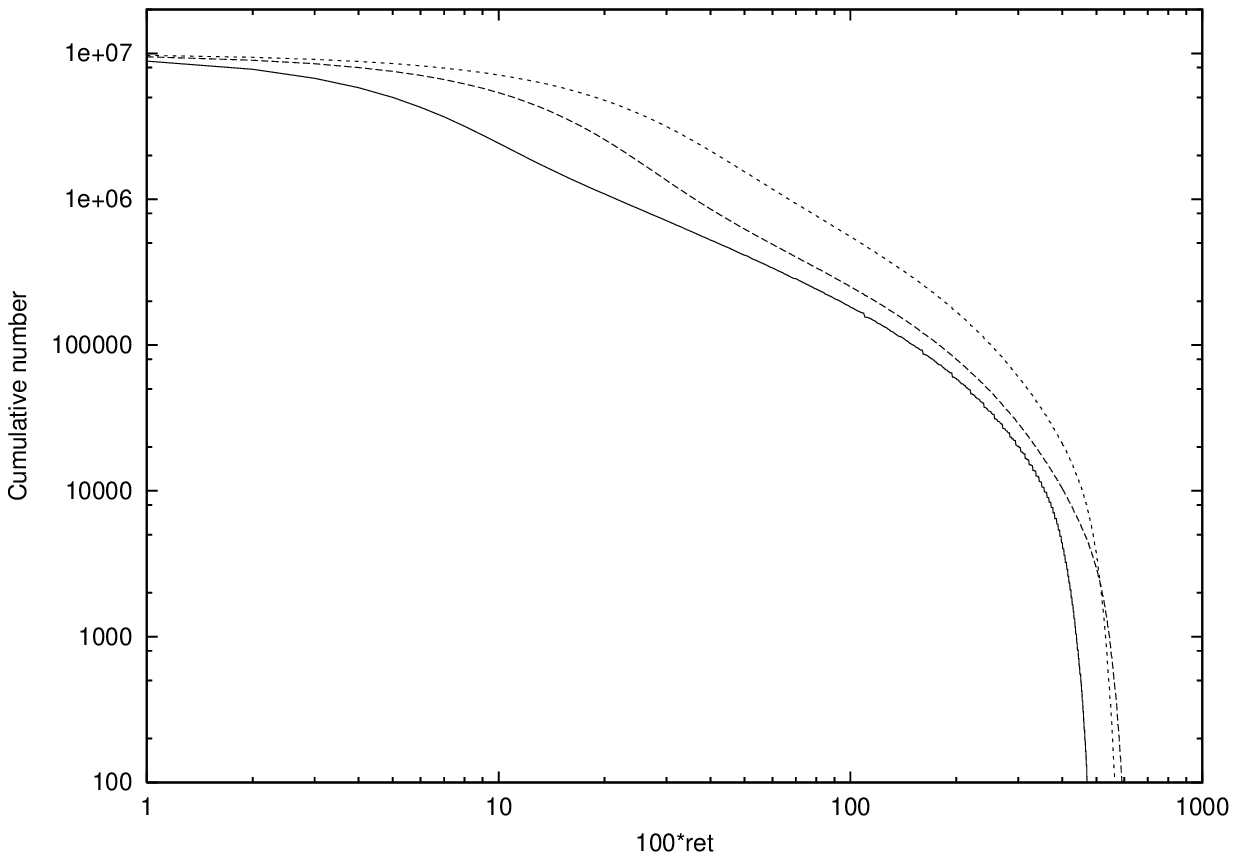}} 
\vspace*{13pt}
\fcaption{Cumulative distribution of absolute returns multiplied by $100$.
For all lines, $J=1.0$,$\alpha =4.0$. The solid,broken and dotted line 
correspond to $2d$ with $T=1.5$, $3d$ with $T=3.8$ and $4d$ with $T=6.0$ 
respectively. For better statistics, we have summed over $10^{7}$ 
Monte Carlo steps.}
\end{figure}

\noindent
\begin{figure}[htbp]
\vspace*{13pt}
\centerline{\psfig{file=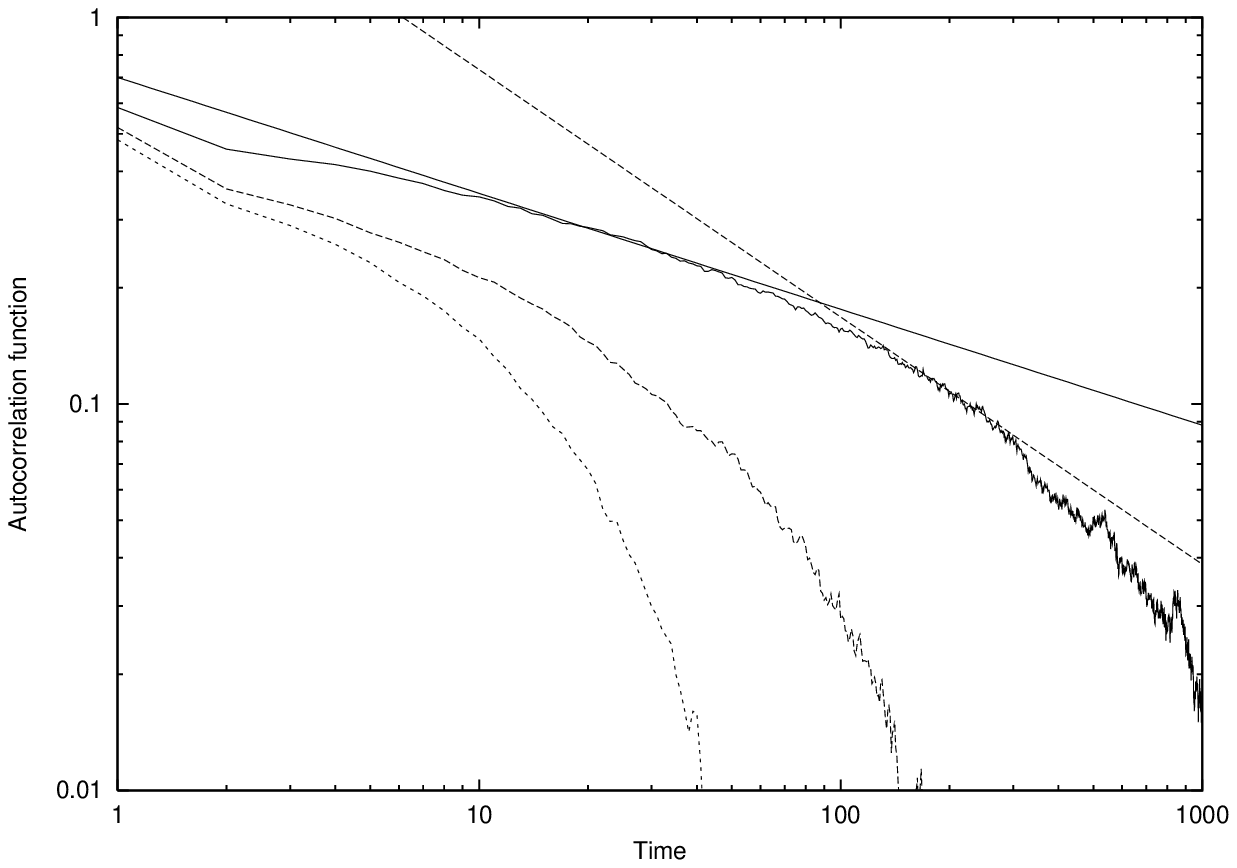}} 
\vspace*{13pt}
\fcaption{Autocorrelation function of absolute returns calculated from 
$10^{6}$ steps. 
The curves correspond to the $4d$, $3d$ and $2d$ from left to right. 
The solid straight line denotes the slope $-0.3$ from references
\cite{Gopi,Liu}. The broken straight line is a slope $-0.64$ from 
a model of minority game\cite{MG}.}
\end{figure}

\noindent
\begin{figure}[htbp]
\vspace*{13pt}
\centerline{\psfig{file=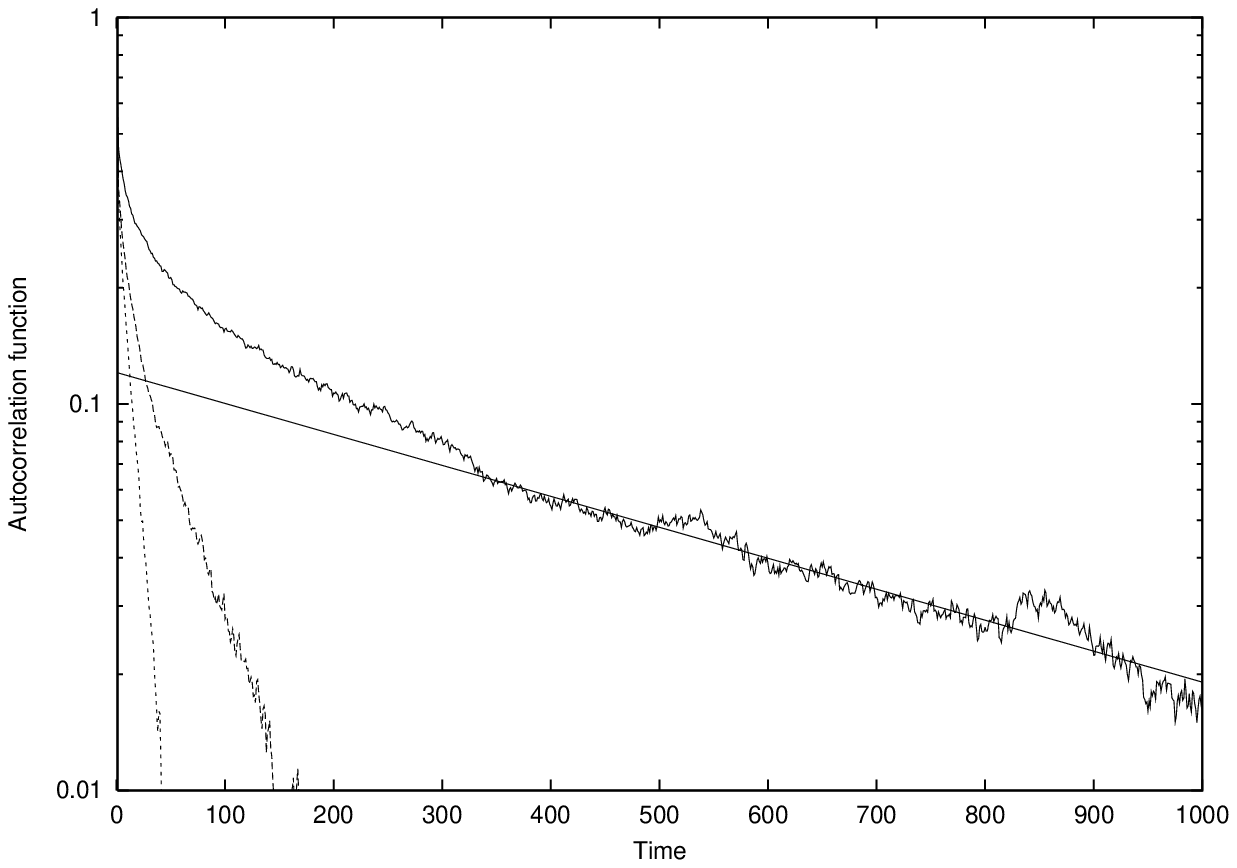}} 
\vspace*{13pt}
\fcaption{Autocorrelation function of absolute returns in semi-log 
scale replotted. The fitted straight line to the $2d$ data has exponent 
$-1.85\times 10^{-3}$.}
\end{figure}

\noindent
\begin{figure}[htbp]
\vspace*{13pt}
\centerline{\psfig{file=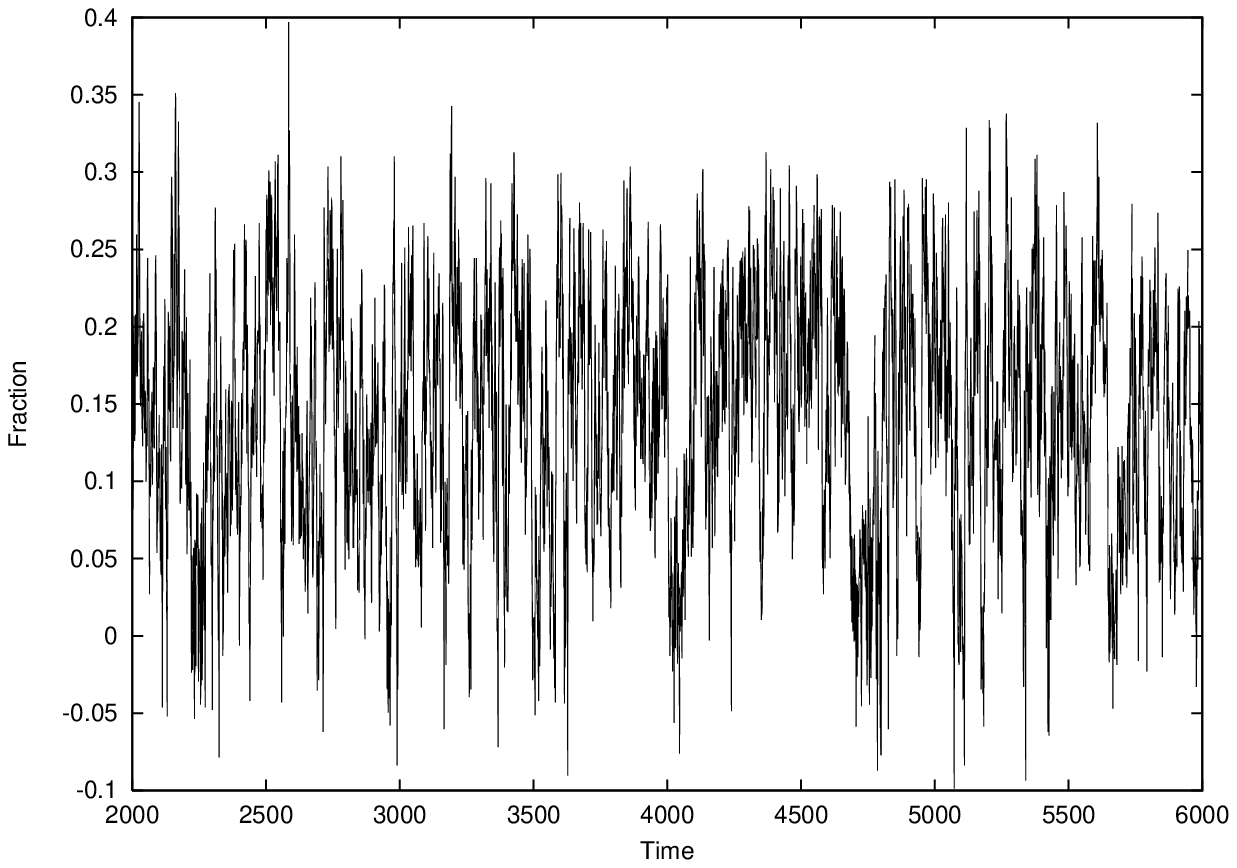}} 
\vspace*{13pt}
\fcaption{The change of the market constituents in $4d$ case ($J=1.0$, 
$\alpha =4.0$ and $T=6.0$) at a period of $4000$.}
\end{figure}

\nonumsection{Acknowledgements}
\noindent
We acknowledge the financial support from
 the DAAD (Deutscher Akademischer Austauschdienst) for staying the
Universit\"{a}t zu K\"{o}ln, where this work was suggested by D. Stauffer.
We also thank S. Bornholdt for his kind explanation of his paper.

\end{document}